\documentclass[psfig,epsfig]{aastex}
\usepackage{emulateapj5}
\usepackage{apjfonts}
\received{}
\accepted{}

\newcommand{\ovi}{{\rm O{\sc vi}\,}}
\newcommand{\ovii}{{\rm O{\sc vii}\,}}
\newcommand{\novi}{{\rm N(O{\sc vi})\,}}

\slugcomment{Submitted to the Astrophysical Journal (Letters), day
May 2005} \lefthead{Avillez \& Breitschwerdt} \righthead{OVI
Distribution in The Galactic Disk}
\begin{document}
\title{Testing Global ISM Models: A Detailed Comparison of O{\sc vi} Column
Densities with FUSE and Copernicus Data}
\author{Miguel A. de Avillez$^{1,2}$ and Dieter Breitschwerdt$^{2}$}
\affil{$^{1}$Department of Mathematics, University of \'Evora, R. Rom\~ao Ramalho 59, 7000 \'Evora, Portugal \\
$^{2}$ Institut f\"ur Astronomie, University of Vienna,T\"urkenschanzstra{\ss}e 17, A-1180 Vienna, Austria\\
E-mail: mavillez,~breitschwerdt@astro.univie.ac.at
}

\begin{abstract}
We study the \ovi distribution in space and time in a representative
section of the Galactic disk by 3D adaptive mesh refinement HD and
MHD simulations of the ISM, including the disk-halo-disk
circulation. The simulations describe a supernova driven ISM on
large ($\sim 10$ kpc) and small ($\sim 1.25$ pc) scales over a
sufficiently large timescale ($\sim 400$ Myrs) in order to establish
a global dynamical equilibrium.
The \ovi column density, \novi, is monitored through lines of sight
measurements at different locations in the simulated disk. One has
been deliberately chosen to be inside of a hot bubble, like our own
Local Bubble, while the other locations are random. We obtain a
correlation between \novi and distance, which is independent of the
observer's vantage point in the disk. In particular, the location of
the observer \emph{inside a hot bubble} does not have any influence
on the correlation, because the contribution of an individual bubble (with a
typical extension of 100 pc) is negligibly small. We find a
remarkable agreement between the \ovi column densities (as a
function of distance) and the averaged \ovi density ($\sim 1.8\times
10^{-8}$ cm$^{-3}$) in the disk from our simulations and the values
observed with \textsc{Copernicus}, and FUSE. Our results strongly
support the important r\^ole of turbulent mixing in the distribution
of \ovi clumps in the ISM. Supernova induced turbulence is quite
strong and unavoidable due to shearing motions in the ISM and
operates on a large range of scales.

\end{abstract}

\section{Introduction}
\label{intro}
Historically, the discovery of a wide-spread diffuse
\ovi component with the \textsc{Copernicus} satellite in absorption
towards background stars, which was not attributable to the
circumstellar environment, led to the establishment of the hot phase
of the ISM (Rogerson et al. 1973; York 1974; Jenkins \& Meloy 1974).
Attempts to link this to the soft X-ray background (SXRB) discovered
earlier (Bowyer et al. 1968) were doomed to fail, because the
excitation temperatures were too different: collisionally ionized
\ovi traces gas of $\sim 3 \times 10^5$ K, whereas the SXRB monitors
$\sim 10^6$ K gas. In collisional ionization equilibrium (CIE) this
difference is impossible to reconcile, since at SXRB temperatures the
ion fraction of \ovii is dominating \ovi by more than two orders of
magnitude. Subsequently, the idea that \ovi was produced mostly in
conductive interfaces between hot and cool gas (McKee \& Ostriker
1977; Cowie et al. 1979) rather than in supernova remnants gained a
lot of attraction.  However, heat conduction in a magnetized plasma is
substantially reduced to the order of a gyroradius, perpendicular to
the magnetic lines of force. If the field lines are highly tangled on
small scales, as it is expected in a turbulent medium, also heat
transfer parallel to the field is severely reduced by effects of
electron trapping in magnetic mirrors. For a detailed discussion see
Malyshkin \& Kulsrud (2001) and Avillez \& Breitschwerdt (2004).

The \ovi distribution in the Galactic disk, within the Local Bubble
(LB) and near the midplane, have been discussed by, e.g., Oegerle et
al.~(2005) and Bowen et al. (2005), respectively. The former
authors, using FUSE data, showed that the \ovi column densities,
\novi, in absorption seen along lines of sight (LOS) against 13, of
the 25 white dwarfs, located within a distance up to 130 pc from the
Sun, are $<10^{13}$ cm$^{-2}$. No \novi~exceeded $1.7\times 10^{13}$
cm$^{-2}$. The average \ovi~number density in the Local ISM found by
these authors is $2.4\times10^{-8}$ cm$^{-3}$, a factor 1.4 larger
than the overall \ovi~density ($1.7\times 10^{-8}$ cm$^{-2}$)
estimated by Bowen et al. (2005), who combined FUSE absorption measurements
against stars located at distances $>1$ kpc with data from
\textsc{Copernicus}. These results indicate that the \ovi has a
clumpy distribution.

In this letter we show that within the framework of the SN-driven ISM
model by Avillez (2000) and Avillez \& Breitschwerdt (2004, 2005;
henceforth AB04 and AB05, respectively), which does not include heat
conduction, both the distribution of \novi~and its dispersion with
distance, as well as the mean \ovi density in the Galactic disk agree
remarkably well with observations. Our simulations emphasize the
importance of turbulent diffusion as an efficient process in mixing
cold and hot gas, an aspect which hitherto has not been studied in
sufficient detail.

\section{Model and Simulations}

In this study we use previously published HD and MHD (with a total
field of 4.5 $\mu$G, with the mean and random components of
$B_{u}=3.1$ and $\delta B=3.2$ $\mu$G, respectively) adaptive mesh
refinement simulations of the SN-driven ISM.  The simulations use a
grid centred at the solar circle with a square disk area of 1
kpc$^{2}$ and extending from $z=-10$ to 10 kpc in the direction
perpendicular to the Galactic midplane. The finest resolution is
1.25 pc. The model includes a gravitational field provided by the
stars in the disk, radiative cooling assuming CIE and solar
abundances (e.g., Sutherland \& Dopita 1993), uniform heating due to
starlight varying with $z$ and a magnetic field (setup at time zero
assuming equipartition) for the case of MHD runs.  SNe types Ia and
II are the sources of mass, momentum and energy. SNe Ia are randomly
distributed, while SNe II have their progenitors generated in a
self-consistent way according to the mass distribution in the
simulated disk and are followed kinematically according to the
velocity dispersion of their progenitors. In these runs we do not
consider heat conduction. For details on the setup and simulations
see AB04 and AB05.

\section{Results}

\begin{figure}[thbp]
\centering
\includegraphics[width=0.3\hsize,angle=0]{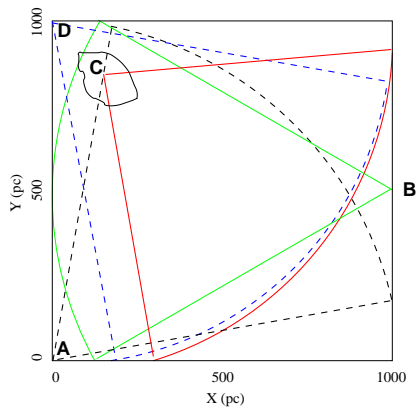}
\caption{ Spatial distribution, in the simulated Galactic midplane
of the locations A, B, C and D, from which the lines of sight are
taken and span a projected area of 90 (B), 85 (C) and 70 (A and D)
degrees. The LOS cross the data cubes up to projected
distances of 1 kpc for all the locations except for position C
(located inside a bubble; see right panel of Figure 2), where the
maximum length is only 800 pc. \label{locations} }
\end{figure}

\begin{figure*}[thbp]
\centering
\includegraphics[width=0.45\hsize,angle=0]{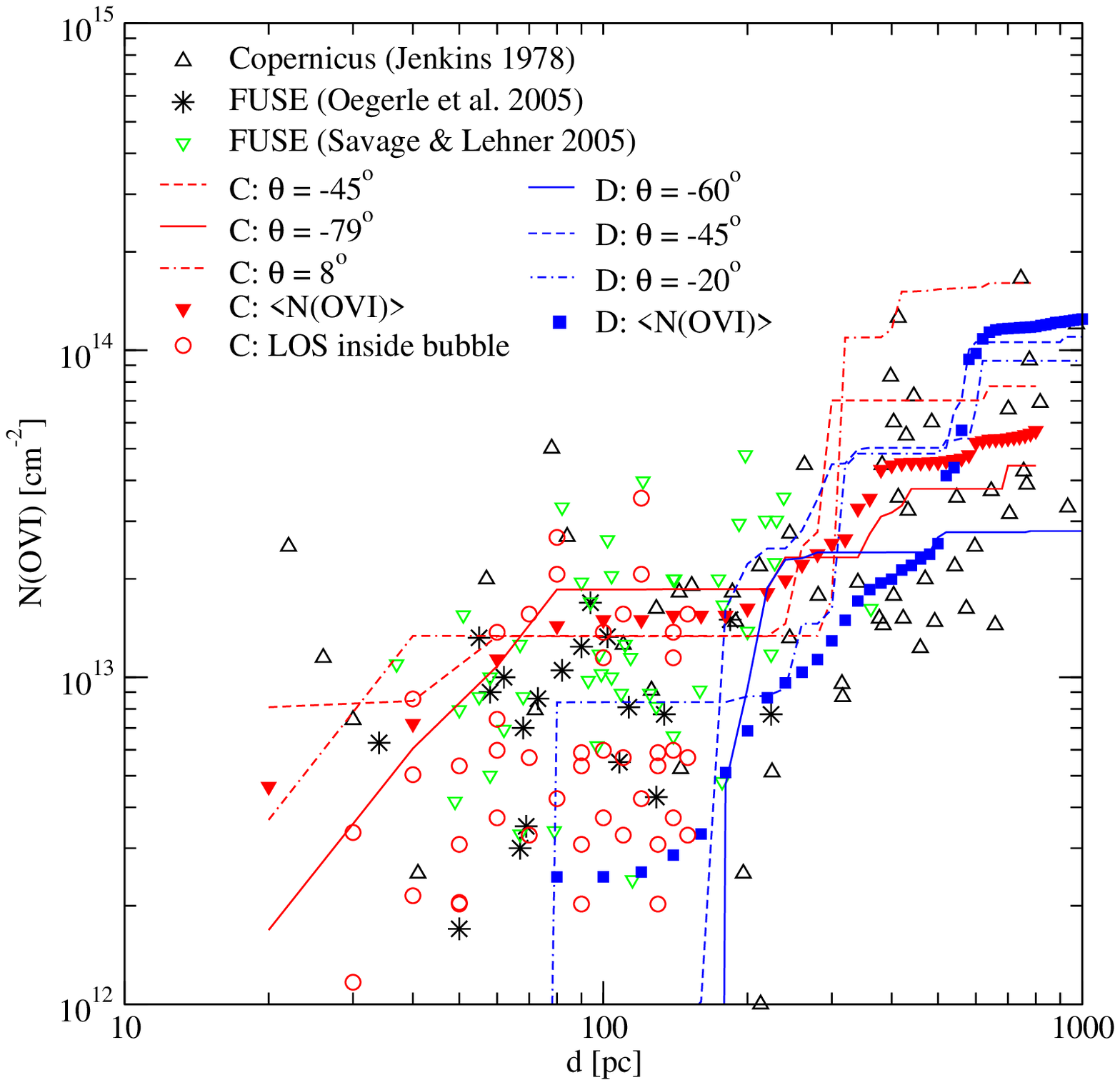}\includegraphics[width=0.55\hsize,angle=0]{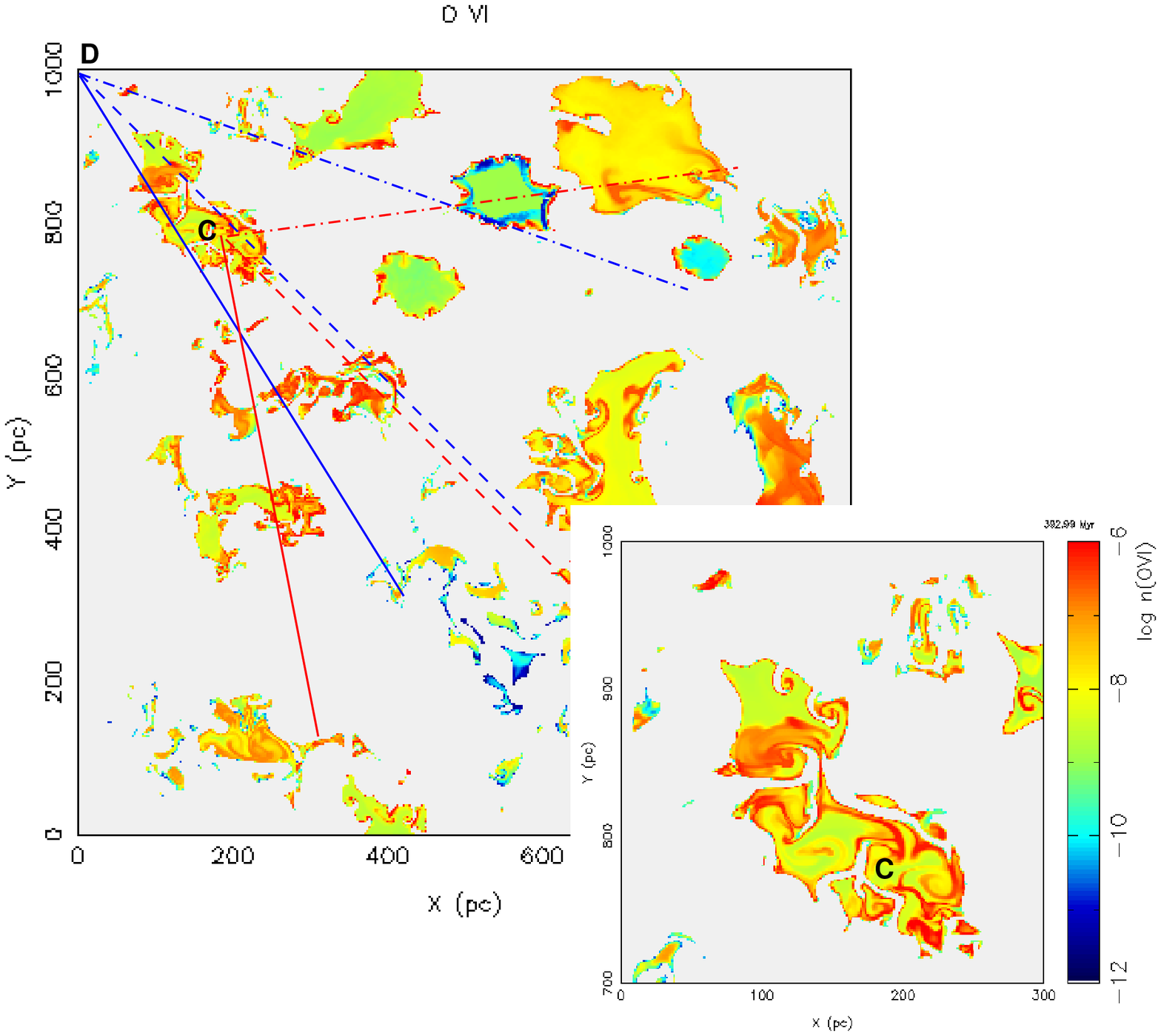}
\caption{\emph{Left panel:} Comparison of FUSE (stars: Oegerle et
al. 2005; green triangles: Savage \& Lehner 2005) and
\textsc{Copernicus} (black triangles) O{\sc vi} column densities with
spatially averaged (red triangles and blue squares) and single lines
of sight (red and blue lines) N(O{\sc vi}) measurements in the
simulated disk at time $t=393$ Myr. The LOS are taken at positions C
(red) and D (blue), which are located inside and outside of a bubble
cavity, respectively, as shown in the right panel. The panel also
shows \novi measurements along LOS sampling gas inside the cavity (red
circles) along directions other than those shown by the red
lines. Note that the variation in \novi inside the bubble is similar
to that observed with FUSE for LOS $<150$ pc. \emph{Right panel:}
O{\sc vi} density distribution (in logarithmic scale) in the midplane
at time $t=393$ Myr. The panel also includes a blow-up of the bubble
located in position C. The colour scale varies between $10^{-12}$ and
$10^{-6}$ cm$^{-3}$; grey corresponds to n(\ovi)$\leq 10^{-18}$
cm$^{-3}$. Note the eddy-like structures of \ovi inside the
bubbles. \label{insidebubble}}
\end{figure*}

Our analysis is restricted to the disk region $\left|z\right|\leq 250$
pc. We took LOS, with $1^{o}$ interval, from different positions
(Figure~\ref{locations}) located in the midplane at coordinates (x=0,
y=0) pc (Pos. A), (x=1000, y=500) pc (Pos. B), (x=200, y=800) pc
(Pos. C) and (x=0, y=1000) pc (Pos. D). Position C is located inside
of a hot bubble similar to our LB (right panel of
Figure~\ref{insidebubble}). At positions A and D we took 71 LOS
covering a region of $70^{o}$, while at positions B and C we took 91
and 86 LOS spanning regions of $90^{o}$ and $85^{o}$,
respectively. The maximum LOS lengths are 1 kpc for positions A, B and
D and 0.8 kpc for position C. The step length is 10 pc. The
periodicity of the boundary conditions along the $x-$ and
$y-$directions assures the continuity with surrounding regions that
are not calculated, thereby not affecting the column densities near
the boundaries.


The left panel of Figure~\ref{insidebubble} compares
\textsc{Copernicus} (black triangles: Jenkins 1978) and FUSE (stars:
Oegerle et al. 2005 and green triangles: Savage \& Lehner 2005) data
with \novi~measurements along LOS taken at positions C (red) and D
(blue) at time $t=393$ Myr. The red circles correspond to \novi
measurements along LOS (with lengths $\leq 150$ pc) that sample gas
inside the bubble centred at position C (right panel of
Figure~\ref{insidebubble}). These measurements show that the \novi
variation inside the cavity is similar to that observed with FUSE in
the LB. The red and blue lines represent \novi along specific LOS
shown in the right panel of Figure~\ref{insidebubble}, while the red
triangles and blue squares correspond to spatially averaged \novi~over
the 86 and 71 LOS taken at positions C and D, respectively (see
Figure~\ref{locations}). These measurements indicate that: (i) for an
observer located at position D (outside of a bubble) there is no
detection of \ovi~until the LOS cross a nearby bubble, (ii) for
$d>100$ pc, i.e.\ a distance much larger than the typical size of a
bubble, the \novi~correlates nicely with distance, and (iii) the \novi
measurements are independent of the observer's location inside of a
hot bubble, because of its small contribution to the total amount of
\ovi~sampled outside of the cavity for $d\gg 100$ pc.

A considerable amount of \ovi is generated in the interiors of
regions created by SNe as a result of local cooling promoted by
turbulent mixing (see right panel of Figure~\ref{insidebubble}, as
well as the blow-up of the bubble in location C, shown in the same
panel). The main advantage of turbulent mixing as compared to
diffusion is the increase of the interaction surfaces between the
cool/warm and hot gas, thus promoting faster cooling. Hence, the
r\^ole of heat conduction (if permitted by a magnetic field) in
these cooling interfaces will be diminished by the reduction of the
temperature gradients due to the turbulent mixing.

\begin{figure}[thbp]
\centering
\includegraphics[width=0.45\hsize,angle=0]{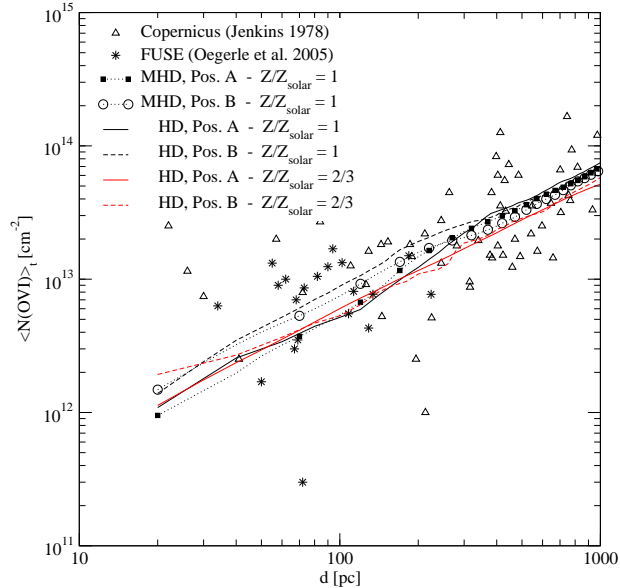}
\caption{
Time averaged O{\sc vi} column densities, $\langle
\mbox{N(OVI)}\rangle_{t}$, as a function of distance for HD and MHD
runs with solar (black lines) and HD run with 2/3 solar (red lines)
abundances, overlayed on FUSE (stars; Oegerle et al.) and
\textsc{Copernicus} (triangles) data. The lines of sight were taken at
positions A and B (see Figure~\ref{locations}). The time average is
calculated over a period of 100 and 50 Myr in the cases of the solar
and subsolar abundances, respectively. \label{comparison} }
\end{figure}

With increasing distance there are jumps (roughly every 100 pc,
indicating that the mean free path between regions with high
\ovi density is of this order) in the column density profiles shown
in the left panel of Figure~\ref{insidebubble} indicating that \ovi
clumps resulting from cooling interfaces are intersected by the
LOS. However, on larger scales (a few hundred pc), due to a repetition
in the ISM pattern distribution of the various phases, the overall
\ovi distribution appears smoother (see Figure~\ref{comparison}).

In order to see that the \novi correlation with distance is not a
transient phenomenon we calculated the time average \ovi column
density, $\langle$\novi$\rangle_{t}$, over the period 301-400 Myr
using 100 data cubes separated by 1 Myr, along the 71 LOS taken at
positions A and B for the HD and MHD runs.  Figure~\ref{comparison}
compares the $\langle$\novi$\rangle_{t}$ in the HD and MHD runs with
FUSE (stars) and \textsc{Copernicus} (triangles) data. The figure
shows a remarkable overlap between the observed and simulated column
densities. In particular, the \novi correlation with distance is
reproduced in the averaging process over a period of 100 Myr of disk
evolution. Thus, such a correlation is independent of time, provided
that the global (not the local) SN rate is constant with
time. Furthermore, it is independent of the observer's vantage point
from which the LOS, sampling gas at distances $>100$ pc, are taken.

\begin{figure}[thbp]
\centering
\includegraphics[width=0.3\hsize,angle=-90]{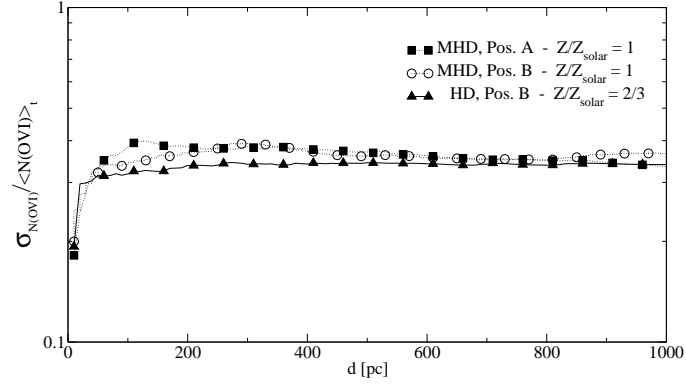}
\caption{
Variation with distance of the \novi dispersion for the measurements
at positions A and B in the MHD (solar abundances) and position B in
the HD (2/3 solar abundance) runs shown in
Figure~\ref{comparison}. \label{dispersion} }
\end{figure}

Although the \ovi~is distributed smoothly enough for \novi~to
correlate with distance, the dispersion $\sigma_{\small{\mbox{N(O{\sc
vi})}}}/\langle\mbox{N(O{\sc vi})}\rangle$ is independent of distance
(Figure~\ref{dispersion}) once $d > 100$ pc.  This result is fully
consistent with observations (e.g., Bowen et al.~2005).

\begin{figure}[thbp]
\centering
\includegraphics[width=0.3\hsize,angle=-90]{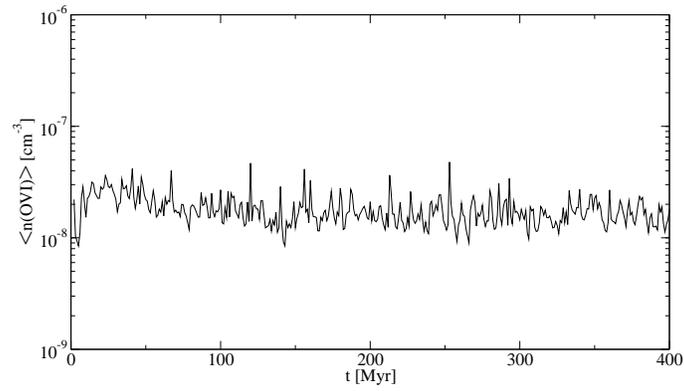}
\caption{
Time evolution of the $\langle\mbox{n(\ovi)}\rangle$ in the simulated
disk (using solar abundances). The mean of the distribution is located
at $\sim 1.8\times 10^{-8}$ cm$^{-3}$.
\label{history} }
\end{figure}

The average of the ratio of the $\langle$\novi$\rangle_{t}$, shown in
Figure~\ref{comparison}, and the distance yields an \ovi~number
density of $1.7-2.1\times10^{-8}$ cm$^{-3}$ for the HD and MHD
cases. This value is in agreement with the time averaged value of
n(\ovi)$\sim 1.81\times 10^{-8}$ cm$^{-3}$ found in the simulated
disks during the period $100 \leq t\leq 400$ Myr
(Figure~\ref{history}). At time 393 Myr $\langle$\novi$\rangle=2\times
10^{-8}$ cm$^{-3}$. The oscillations in n(\ovi) around the mean value
seen in the figure are correlated with local variations in the SN
rate. In fact, taking into account that the number of stars during the
simulations at any time are determined self-consistently from the
amount of cold gas present in the disk at that time, it is not
difficult to conclude that spatial variations in the star formation
rate occur, generating local variations in the \ovi density.

\section{Discussion}
The picture that emerges from the present simulations is that \ovi
absorption arises from clumps distributed in the highly turbulent
interstellar medium. As shown in this letter, without taking into
account heat conduction (which will be the subject of a forthcoming
paper), the simulations reproduce the \ovi distribution that has
been measured with FUSE and \textsc{Copernicus}. The clumpy
distribution is a natural consequence of turbulent diffusion,
induced by shear motions, that efficiently transports and mixes cold
and hot gas inside the bubbles and at small scale interfaces. Shear
motions destroy large scale surfaces of adjacent cold and hot gas by
generating vorticity, which stretches the fluid into thin filaments.
Therefore, the patchiness of the \ovi distribution is retained, but
the amount of \ovi per interface is substantially reduced.

Oegerle et al. have argued that heat conduction across
interfaces between clouds and hot gas generates the observed
\ovi~distribution. But, since the process is too efficient it has to
be quenched significantly by invoking tangled magnetic fields on the
clouds' surfaces, resulting in a patchy distribution. Although this
mechanism can in principle operate, it seems to be contrived as the
quenching factors have to be assumed ad hoc to fit the data.  Due to
the lack of predictive power of conduction, we favour turbulent
mixing, which arises naturally from a SN driven ISM, as the main
process for generating \ovi patches in the ISM, also explaining the
data without further tuning of the model. In fact, heat conduction,
depending on local temperature gradients, is usually slower than
turbulent mixing. Still heat conduction can be important locally, if
the magnetic field topology permits. In a SN driven ISM, on the other
hand, turbulent mixing will be inescapable, independent of the field
geometry, as vortex stretching implies field line stretching down to
scales where tension forces become strong enough to counteract.

The correlation of \novi with distance for $d> 100$ pc indicates
that the processes, which give rise to the existence of hot gas in
the Galactic disk are ubiquitous in all the simulated ISM in both HD
and MHD runs, and ISM patterns repeat on scales of a few 100 pc. The
small contribution that hot bubbles have to the \ovi distribution
(and \novi correlation with distance) seen in LOS sampling gas
\emph{outside} the bubbles itself, indicates that the Local Bubble has a
negligible contribution to the \novi measured for LOS distances
$>120$ pc.

It has been argued recently that the abundances in the local ISM are
roughly 2/3 solar (e.g., Meyer 2001) hence, -log(O/H)=-3.46 instead of
-3.07 for solar abundances (Anders \& Grevesse 1989). We have
therefore repeated the HD run (but for a time evolution of 250 Myr due
to computing time limitations) for this case with the result of slight
reductions in the \ovi dispersion (black triangles in
Figure~\ref{dispersion}) and in \novi (see Figure~\ref{comparison},
where the red lines show time averaged, over the period 200-250 Myr,
\novi measured at positions A and B). These findings can be understood
if one realizes that a reduced metallicity also leads to a reduced cooling
of the interstellar gas, and hence to larger cooling times.
Using CIE cooling curves, the
transition in the temperature range where \ovi is the dominant
ionization stage takes longer compared to the solar abundance case.
Therefore we are likely to observe a somewhat higher amount of \ovi in
the supernova driven ISM, compensating to some extent the lower oxygen
abundance. Our calculations do not include depletion of oxygen onto
dust grains. This might reduce \novi somewhat, although we do not
expect depletion in this temperature range to be very high.

\section{Concluding Remarks}

It should be emphasized that no ``tuning'' of the simulations was
applied. Both HD and MHD runs and the resulting data cubes were
obtained \emph{before} we had access to the FUSE data. There are no
free physical parameters in the set-up (we fixed whatever we could to
observed Galactic values, e.g.\ diffuse photon field, gravitational
field etc.). All we used was a starting model that corresponded to the
\emph{currently} observed H{\sc i} and H{\sc ii} distribution in the
Galactic disk, and a SN rate fixed initially at the current rate. Then
we let the system evolve 
for a sufficiently long time.  Therefore, either the
system is very \emph{insensitive} to physical boundary conditions, or,
we have captured the major features of the ISM, resulting in a global
dynamical equilibrium. Although it is true that the system is
self-regulating to some extent, the first possibility could not be
confirmed, since volume filling factors depend sensitively on the SN
rate (especially for the hot gas) and the establishment of a Galactic
fountain flow.  We therefore favour the second possibility, with a
corollary, that conductive interfaces, which should be abundant and
exhibit a significant amount of \ovi, if heat conduction is a major
transport process in the ISM, are unlikely to dominate the
redistribution of energy. This task is more efficiently done by
turbulent mixing. A detailed study of the results discussed here, as
well as on the kinematics of the \ovi in the ISM, will be presented in
a forthcoming paper.

This research is supported by FCT through grant BSAB-455 to M.A. and
project PESO/P/PRO/40149/2000 to the authors. We thank the two
anonymous referees for helpfull criticisms.

\end{document}